\documentclass[prb,twocolumn,showpacs]{revtex4}

\usepackage{bm}
\usepackage[dvips]{graphicx}
\usepackage{color}
\usepackage{latexsym}
\usepackage{amsmath}
\usepackage{amssymb}
\usepackage{psfrag}

\begin{document}

\preprint{---}


\title{Classical magnetotransport of inhomogeneous conductors}

\author{Meera M. Parish}
  \email{mmp24@cam.ac.uk}
\author{Peter B. Littlewood}

\affiliation{Cavendish Laboratory, Madingley Road, 
  Cambridge CB3 0HE, United Kingdom}

\date{\today}

\begin{abstract}
We present a model of magnetotransport of inhomogeneous conductors 
based on an array of coupled four-terminal elements.  We show that this 
model generically yields non-saturating magnetoresistance
at large fields.  We also discuss how this approach simplifies 
finite-element analysis of bulk inhomogeneous semiconductors in 
complex geometries.  We argue that this is an explanation of 
the observed non-saturating magnetoresistance in silver chalcogenides 
and potentially in other disordered conductors.  Our method may be 
used to design the magnetoresistive response of a microfabricated 
array.
\end{abstract}

        
\maketitle
%
%
\section{Introduction}

The problem of determining the effective conductivity of a classical 
inhomogeneous medium~\cite{I92} is an old one, but a comprehensive theory 
of the magnetotransport in such systems is still far from complete.
Moreover, much of the early work on inhomogeneous conductors concentrates 
on the zero-magnetic-field case since this is analogous to calculating the 
polarization of a random dielectric \cite{summary}.
For the classical problem to be appropriate, the mean free path of the charge 
carriers must be much less than the typical length scale of the disorder 
so that Ohm's law is obeyed locally in space:
\begin{equation}
    {\bf E}({\bf r}) \ = \ \hat\rho({\bf r}) \ {\bf j}({\bf r})
\end{equation}
where ${\bf j}$ is the current density, ${\bf E}$ is the electric field and 
$\hat\rho$ is the resistivity tensor.  In the case of a simple conductor 
that possesses a single charge carrier and isotropic inhomogeneities, the 
resistivity tensor acquires the following form in a magnetic field $H$:
%
\begin{equation}\label{rho}
\hat\rho \ = \ \rho_0
\left (
\begin{array}{ccc}
1 & \beta & 0 \\ 
-\beta & 1 & 0 \\
 0 & 0 & 1
\end{array}
\right )
\end{equation}
It is clear that this tensor is characterised by just two parameters: 
the carrier mobility $\mu$ (since the dimensionless variable 
$\beta \equiv \mu H$) and the scalar resistivity $\rho_0$. 
While more complex, anisotropic resistivity tensors have also been 
considered in the inhomogeneous conductor problem~\cite{dreizin,IK91,bala8},
this paper shall be solely concerned with Eq.~(\ref{rho}).

For inhomogeneous conductors, the effective resistivity 
$\rho_{\mathrm{eff}}$ is defined by 
$\left< \bf E \right>_v = \rho_{\mathrm{eff}} \left< \bf j \right>_v$,
where $\left< \cdots \right>_v$ specifies an average over volume.
In the absence of disorder, the magnetoresistance $\Delta R/R \equiv 
[\rho_{\mathrm{eff}}(H)-\rho_{\mathrm{eff}}(0)]/\rho_{\mathrm{eff}}(0)$ 
is trivially zero, for arbitrary orientation of the magnetic field.
However, in general, the magnetoresistance strongly depends on whether 
it is transverse or longitudinal, i.e.\ whether the magnetic field is 
perpendicular or parallel to the current. Furthermore, any nonzero 
transverse magnetoresistance must be an even function of field due to the 
rotational symmetry about the current axis.  Thus, $\Delta R/R \propto H^2$ 
in the low field limit, which is usually defined to be $\beta \ll 1$, but 
the crossover from the low-field to high-field regime is not always 
obvious in disordered systems, as will be discussed later. 

Some recent experiments~\cite{silver1, silver2, silver3,
SSR00,OHB99,LCWT00,CD98,MPKS01,KMGKKJ05,BKJGK04} 
on the doped silver chalcogenides, Ag${}_{2+\delta}$Se and Ag${}_{2+\delta}$Te,
have added impetus to the investigation of this problem.  Both silver 
chalcogenides exhibit a positive, transverse magnetoresistance that is a 
linear function of magnetic field 
throughout the temperature range $4.5$K to $300$K, with no signs of 
saturation up to fields of 60T\cite{silver1, silver2}. 
In particular, the linearity continues down deep into 
the low-field regime $\beta \ll 1$.
Such behaviour is not what is seen in conventional semiconductors, 
where the resistance increases quadratically with increasing 
magnetic field at low fields and, except in very special circumstances,
eventually saturates at fields typically of order 1T \cite{kittel,semi}, 
corresponding to $\beta \sim 1$.
Since the silver chalcogenides are non-magnetic compounds, 
the origin of the large magnetoresistance is 
unclear, although a quantum theory based on the partial population of one 
Landau magnetic band has been proposed \cite{abri, abri2}.  
However, the large range in temperature over which the phenomenon 
occurs suggests that one should examine large magnetoresistances resulting 
from classical effects, namely the case where the semiconductor is highly 
inhomogeneous.  

The theoretical study of classically disordered conductors 
may be divided into two separate classes: 
media consisting of two or more distinct phases, separated 
by sharp boundaries, and systems that possess continuously variable 
fluctuations in the conductivity.  In the first class, 
solutions for a non-zero magnetic field have been derived for 
an isotropic medium with a low volume 
fraction $c \ll 1$ of insulating spherical inclusions \cite{bala, stroud}, 
and they give a positive linear magnetoresistance in the high field limit 
$\beta \gg 1$, but the increase of the magnetoresistance 
$\Delta R/R$ with field is small, being proportional to $c$.
An effective medium method has been used to extend this solution to 
higher volume fractions \cite{BS00}, but this result is approximate and 
it is still only applicable to high fields $\beta\gg 1$.
The class of systems with continuously varying conductivity fluctuations 
has only been studied for weak, short-range disorder \cite{herring, dreizin}. 
Using an advection-diffusion analogy, the effective magnetoresistance is 
determined to be $\Delta R/R \sim \gamma^{4/3} \beta^{2/3}$ for 
$\beta \gamma \gg 1$, where the disorder width $\gamma \ll 1$.

Thus, the main limitation of the current literature is that it is generally 
restricted to media that only deviate slightly from homogeneity, so the 
increase in magnetoresistance is small and anomalous behaviour only occurs at 
very high magnetic fields.  Whilst there is an exact solution for the 
effective magnetoresistance in two dimensions that yields a linear
magnetoresistance, it is restricted to the special case of a two-component 
media with equal proportions of each phase \cite{D71,B78,GS05}.
However, this does lend credence to the hypothesis that classical disorder 
is the cause of the anomalous magnetoresistance of the silver chalcogenides.

In order to attack the problem of strong inhomogeneities, we previously 
introduced a two-dimensional random resistor network model \cite{me}. 
We used it to show that classical disorder is a possible 
cause of the anomalous magnetoresistance of the silver chalcogenides, 
and we raised the possibility of using the networks to construct magnetic 
field sensors that operate on principles similar to extraordinary 
magnetoresistance (EMR) devices \cite{STHH00,ZHS01-2}.

In this paper, we investigate the galvanomagnetic properties of the 
random resistor network model in detail.  
This model allows one to study the magnetoresistance of an inhomogeneous 
semiconductor across the whole magnetic field range for a variety of 
disorder.  By considering voltages and 
current paths within the network, as well as network magnetoresistances, 
we demonstrate that our resistor network is also capable of simulating 
macroscopic media with complicated boundaries.  We use this result to 
include contact effects between resistors within the network.

The paper is organised as follows:
Section~\ref{model} describes the random resistor network model and derives 
expressions for the network magnetoresistance and Hall resistance.  
In Sec.~\ref{small}, we use the insight gained from studying the 
characteristics of small networks to ascertain the symmetries of the 
network magnetoresistance and, thus, determine 
the condition for which the magnetoresistance is non-saturating. 
Next, we examine larger networks by studying the magnetotransport of 
uniform square networks and random square networks in Sec.~\ref{large}.
Finally, we address the ramifications of contact resistances between 
resistors and boundary effects within the resistor network in 
Sec.~\ref{discuss}, before concluding in Sec.~\ref{conc}.

%
%
\section{Resistor network model}\label{model}

We tackle the inhomogeneous conductor problem by discretizing the 
medium into a random resistor network and analysing it numerically. 
Standard resistor networks, where the network unit is taken to be a 
two-terminal homogeneous resistor, are inadequate for simulating current 
flow in a magnetic field, since it does not allow the current to 
flow perpendicular to the voltage drop across a resistor.  
Thus, a network of two-terminal resistors will not faithfully represent 
Eq.~(\ref{rho}), which requires the local current to make an angle 
$\arctan(\beta)$ with the local electric field.

\begin{figure}[htbp]
\begin{center}
\includegraphics[width=3cm]{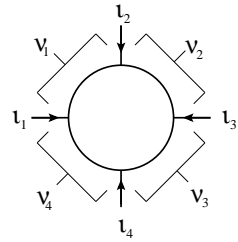}
\caption{The network resistor unit consists of a homogeneous, conducting 
  disk with four equally spaced terminals.  Currents $\iota$ entering the 
  disk are taken to be positive, while the voltage differences $\nu$ 
  between the terminals are considered positive when measured in the 
  clockwise direction.}
\label{res}
\end{center}
\end{figure}

The simplest resistor network model that takes account of the Hall component 
is a two-dimensional square lattice constructed of four-terminal resistors, 
with a magnetic field applied perpendicular to the network. 
This is sufficient for simulating a transverse magnetoresistance, but not 
a longitudinal one since this requires a three-dimensional network. 
Networks with multi-terminal elements have also been used to study 
percolating media in a magnetic field \cite{SBS93} but, unlike our model, 
the elements were restricted to being either insulators or conductors of 
a set conductivity.

In principle, the network resistor unit can be of arbitrary geometry but, 
for simplicity, we take it to be a homogeneous circular disk with four 
current terminals and four voltage differences between the terminals, 
as shown in Fig.~\ref{res}.
These currents and voltages are related via a $4\times 4$ matrix $z$:
\begin{equation}
\nu_{i} \ = \ z_{ij} \ \iota_{j}
\end{equation}
The coefficients $z_{ij}$ can be determined by solving the Laplace equation 
for the electric potential of a homogeneous, conducting disk, using the 
currents as boundary conditions (see Appendix~\ref{appA}).  Note that this 
formulation implicitly assumes a \emph{uniform} injection of current into 
the terminals at all magnetic fields.  In practice, a magnetic field generally
perturbs the current at a boundary between two different conductors, 
but we shall neglect these corrections for now, and revisit 
them later in our discussion of boundary effects in Sec.~\ref{discuss}.

If the terminals are taken to be equally spaced and the angular width 
$\varphi$ of the terminal is held fixed (we will take $\varphi = 0.14$ 
radians in this paper), then the resistor impedance matrix has the form:
\begin{equation}\label{z_disk}
 z = \frac{\rho}{\pi t}
     \left (
     \begin{array}{cccc}
      a & b & c & d \\  
      d & a & b & c \\ 
      c & d & a & b \\ 
      b & c & d & a
     \end{array}
     \right )
\end{equation}
Here, $\rho$ is the disk scalar resistivity and $t$ is the disk thickness, 
while the matrix elements are dependent on $\varphi$ and $\beta$: 
$a=-g(\varphi)+\frac{\pi}{4}\beta$, $b=g(\varphi)+\frac{\pi}{4}\beta$, 
$c=0.35-\frac{\pi}{4}\beta$ and $d=-0.35-\frac{\pi}{4}\beta$.  In the limit 
$\varphi\rightarrow 0$, the function $g(\varphi)\rightarrow \infty$ 
so that the disk resistance diverges as expected.   
Like Eq.~(\ref{rho}), the impedance matrix $z$ of each resistor is 
characterized by two independent parameters: the mobility $\mu$ and 
the quantity $s = \rho/(\pi t)$.
Note that the cyclical permutation of matrix elements is associated with any 
$n$-terminal resistor that is invariant under rotations of $2\pi/n$ radians.  
There is also the added constraint that $\sum_i \nu_i = 0$, so we have 
$a + b + c + d =0$.

\begin{figure}[htbp]
\begin{center}
\includegraphics[width=7cm]{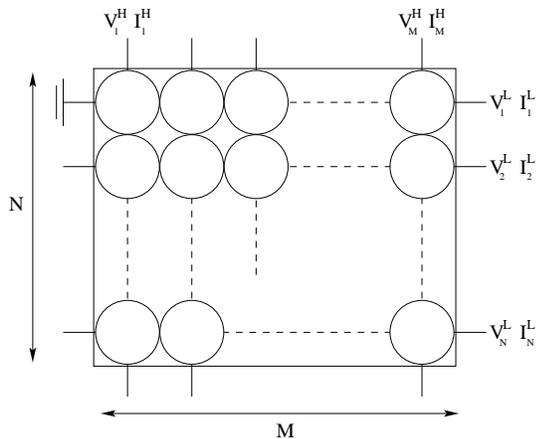}
\caption{Schematic diagram of an $N\times M$ network.  One terminal
  is grounded to provide a point of reference for the voltages, and we 
  can disregard the current at the grounded terminal by imposing current 
  conservation.  Voltages and currents can be classified into $2N-1$ 
  longitudinal components $V^{L}_{i}$, $I^{L}_{i}$ and $2M$ Hall components 
  $V^{H}_{i}$, $I^{H}_{i}$.}
\label{resnet}
\end{center}
\end{figure}

To construct an $N\times M$ random resistor network, we connect the 
disks together (e.g. using perfectly conducting wires) and then vary 
$\mu$ and $s$ for each resistor.  Note that we can include positive 
charge carriers (holes) by allowing $\mu$ to be positive as well as 
negative, whereas previous studies of inhomogeneous media have generally 
focused on charge carriers of the same sign.
In the context of real materials such as the silver chalcogenides, we can 
view the network as representing an array of silver-ion clusters and voids 
within the semiconductor.

A typical $N\times M$ network is depicted in Fig.~\ref{resnet}.
One can define a network impedance matrix $Z$ so that the input 
voltages $V_{i} = Z_{ij}I_{j}$, where $I$ corresponds to the input 
currents.  The impedance matrix is determined by grounding one terminal 
to provide a point of reference for the voltages, and then using Kirchoff's 
laws to eliminate the current at the grounded terminal as well as eliminating 
the internal currents and voltages within the network.
By classifying the voltages and currents into $2N-1$ longitudinal components 
$V^{L}_{i}$,~$I^{L}_{i}$ and $2M$ Hall components $V^{H}_{i}$,~$I^{H}_{i}$, 
the $(2M+2N-1)\times (2M+2N-1)$ impedance matrix $Z$ can be written as
\begin{equation}\label{Zmat}
Z \ = \
\left (
\begin{array}{cc}
   Z^{HH}
   &
   Z^{HL}
   \\
   Z^{LH}
   &
   Z^{LL}
\end{array} \right ) \
\end{equation}
To determine the magnetoresistance of an $N\times M$ network, 
we set $I^{H}_{i}=0$ and 
completely ground the left side of the longitudinal voltages in 
Fig.~\ref{resnet} while setting $V^{L}$ on the right side to a constant 
potential $U$.  The network resistance $R_{NM}(H)$ is then given by:
\begin{equation}\label{netmag}
  R_{NM}(H) \ = \ \frac{U}{\sum_{i}^N I^{L}_{i}}
          \ = \ \frac{U}{\sum_{i}^N 
        \left(Z^{LL}\right)_{ij}^{-1}V^{L}_{j}}
\end{equation}
where the sum over input currents is performed along the ungrounded (right) 
edge.  Similarly, the Hall voltages are
\begin{equation}
V^{H}=Z^{HL}\left(Z^{LL}\right)^{-1}V^{L}
\end{equation}
If we keep the ratio $N/M$ constant and take the limit where 
$N\rightarrow\infty$, then the resistor network should give us the 
galvanomagnetic properties of a real material.
Equation (\ref{netmag}) is difficult to solve analytically for large networks 
and, in practice, we just use Kirchoff's laws to numerically solve for all the 
currents and voltages in the network, since this allows us to study the 
current flow and voltage landscape within a network.  However, considerable 
insight can be gained from examining the symmetries of $Z$.

%
%
\section{Small networks and Network symmetries}\label{small}
In order to elucidate the basic properties of the resistor network model, 
we begin by studying small networks, 
namely $1\times M $ and $N\times 1$ networks. 
\begin{figure}[htbp]
\begin{center}
\includegraphics[width=5cm]{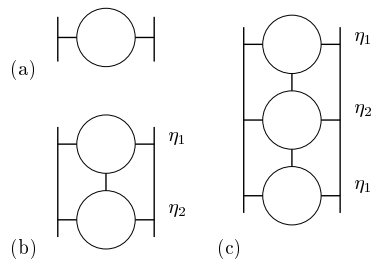}
\caption{Examples of small resistor networks corresponding to 
 (a) a single resistor, (b) $2\times 1$ network and (c) $3\times 1$ network,
  where $\eta_i \equiv (s_i,\mu_i)$.}
\label{smallnet}
\end{center}
\end{figure}
The simplest network is a single resistor, shown in Fig.~\ref{smallnet}(a), 
and this yields $\Delta R/R = 0$ as expected, because we have assumed that 
the disk is a conventional conductor with no implicit magnetoresistance.
Moreover, this behaviour holds for generic $1\times M$ networks which 
are simply equivalent to chains of two-terminal resistors.
However, $N\times 1$ networks exhibit non-trivial behaviour 
since they allow for a plurality of current paths within the network 
when $N>1$.  A $2\times 1$ network of identical resistors 
($\eta_1 = \eta_2$ in Fig.~\ref{smallnet}(b)) yields 
$\Delta R/R \propto \beta^2$ while a $3\times 1$ network of identical resistors
gives a non-zero magnetoresistance that saturates when $\beta \gg 1$.
\begin{figure}[htbp]
\begin{center}
\includegraphics[width=7cm]{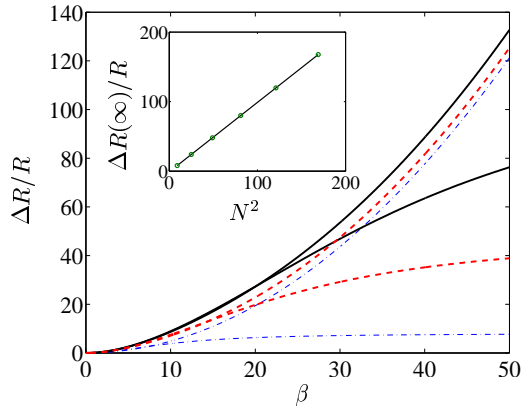}
\caption{(Color online) 
  Magnetoresistance of uniform $N\times 1$ networks with set $\mu, s$, 
  where the saturating curves correspond to $N =$ 3, 7 and 11, in order of 
  increasing size.  
  The remaining curves represent $N=$ 2, 6 and 10, in order of size again.
  Inset: for odd $N$ the saturation level $\Delta R(\infty)/R \simeq N^2$}
\label{Nby1}
\end{center}
\end{figure}
Figure~\ref{Nby1} reveals that among uniform $N\times 1$ networks there is a 
general trend for even-$N$ networks to have a non-saturating magnetoresistance
and for odd-$N$ networks to possess a saturating magnetoresistance, where the 
saturation level $\Delta R(\infty)/R$ scales as $N^2$.
We see that the differences in $\Delta R/R$ between odd-$N$ 
and even-$N$ networks diminish as $N$ increases, with $N$ and $N+1$ 
lying on the same curve for sufficiently small $\beta$. 
However, note that the $N\times 1$ network does not represent a 
well-defined system in the infinite-$N$ limit since $R_{N1}(0)\rightarrow 0$.


%
%

Studies of $N\times 1$ networks consisting of two types of resistors have 
also yielded interesting patterns in the high field behaviour of 
$\Delta R/R$.  For the $2\times 1$ network shown in 
Fig.~\ref{smallnet}(b), the magnetoresistance saturates when 
$s_1\mu_1\neq s_2\mu_2$, where the saturation value of $\Delta R/R$ 
depends on $(s_1\mu_1-s_2\mu_2)$, but setting $\mu_1\neq\mu_2$ and 
$s_1\mu_1 = s_2\mu_2$ always gives a non-saturating magnetoresistance.  
Another example is the $3\times 1$ network in Fig.~\ref{smallnet}(c) which 
always has a saturating magnetoresistance except when $s_1\mu_1=2s_2\mu_2$.  
Similar patterns occur in networks of larger $N$.

The emergence of these symmetries can be understood by considering the 
impedance matrix $Z$. One can demonstrate that it has the form:
\begin{equation}\label{ZSA}
  Z \ = \ S \ + \ \beta A
\end{equation}
where $S$ and $A$ are symmetric and antisymmetric matrices, respectively.  
For chains of resistors, including the case where we only have one resistor, 
$S$ and $A$ are always independent of $\beta$, but in general they are only 
constant in the limits where $\beta \rightarrow 0$ and 
$\beta \rightarrow \infty$.  Note that $Z$ must always be a symmetric matrix 
at zero magnetic field, since an antisymmetric matrix implies dissipationless 
current flow.  To see this, consider the power of the network
\begin{eqnarray} \nonumber
  P & = & V^TI = I^TV \\ \nonumber
    & = & I^T Z^T I = I^T Z I
\end{eqnarray}
Thus, we have $P=0$ if $Z^T=-Z$.  Since the presence of a magnetic field 
induces dissipationless flow, it makes physical sense to have an 
antisymmetric matrix attached to $\beta$. 
%
%

The relevant quantity that is used to derive the magnetoresistance 
is the $(2N-1)\times (2N-1)$ matrix $Z^{LL}$.  From Eq.~(\ref{Zmat}), 
it also has the form $Z^{LL} = S^{LL} + \beta A^{LL}$.
We can write $Z^{LL}=\beta Z_{\beta}$ so that $Z_{\beta}\rightarrow A^{LL}$ 
as $\beta\rightarrow \infty$.
Moreover, $A^{LL}$ is an odd antisymmetric matrix for all $N$, 
so $Z_{\beta}$ will possess at least one eigenvalue that 
approaches zero at large fields.

Now, the sum of the input currents along the right edge can be written as 
\begin{equation}
  \sum^N_i I^L_i \ = \ \frac{1}{\beta} \sum^N_i \sum^{2N-1}_n 
                 \frac{w_{n,i} \ w^T_{n,j}}{\lambda_n} \ V^L_j
\end{equation}
where $w_n$ and $\lambda_n$ are the $n$th eigenvector and eigenvalue of 
$Z_{\beta}$, respectively.  Since $\beta$ appears in the denominator, 
all terms will vanish in the high field limit except for the singular 
terms associated with the eigenvalues approaching zero.  
Their behaviour will determine whether the magnetoresistance is 
saturating or non-saturating.
If we assume that only one eigenvalue $\lambda_0$ approaches zero in the 
high field limit, then we have
\begin{align}\
    \sum^N_i I^L_i \ \simeq \ &  \frac{U}{\beta \lambda_0}
    \left(\sum^N_i w_{0,i}\right)^2 \ + \ O \left(\frac{1}{\beta} \right)
\end{align}
The behaviour of $\lambda_0$ is governed by $S^{LL}/\beta$ so we must have 
$\lambda_0 \propto 1/\beta$ when $\beta \gg 1$.  Therefore, $\lambda_0$ 
 cancels $\beta$ in the denominator and the magnetoresistance is only  
non-saturating if $\sum^N_i w_{0,i} \rightarrow 0$ at large fields.  This 
explains why special configurations of resistors in the small networks give 
a non-saturating magnetoresistance.  In the case where the magnetoresistance 
is saturating, $\lambda_0$ dominates the electrical transport and 
$I^L_i \rightarrow w_{0,i}$.
More generally, $\sum^N_i w_{0,i}$ will determine the exact dependence of 
$R_{NN}(H)$ on field as $\beta \rightarrow \infty$.

To be more concrete, let us consider the simple case where
\begin{equation}
  A^{LL} \ = \ 
\left (
\begin{array}{cc}
   \mathbf{0}
   &
   \mathbf{1}
   \\
   \mathbf{-1}
   &
   \mathbf{0}
\end{array} \right ) \
\end{equation}
Then the zero eigenvalue has the (normalized) eigenvector 
$w_{0i} = (-1)^i/\sqrt{2N-1}$ and in the high field limit we obtain
\begin{equation}
  \sum^N_i I^{L}_i \ \propto \ \frac{U}{2N-1}
   \left[ \left(\sum^N_i(-1)^i \right)^2 \ + \ \delta(\beta,N)\right]
\end{equation}
where $\delta(\beta,N)$ is a finite-field correction factor 
that vanishes as $\beta\rightarrow \infty$.
Therefore we have the high-field resistance 
\begin{equation}\label{univ}
  R_{NN}(H) \ \propto \ 
\begin{cases}
   2N-1
   &
   \text{if $N$ is odd}
   \\
   \frac{2N-1}{\delta(\beta,N)}
   &
   \text{if $N$ is even}
\end{cases}
\end{equation}

The magnetoresistance is non-saturating for even-$N$ networks and 
it saturates for odd-$N$ networks, in agreement with Fig.~\ref{Nby1}.
The fact that $R_{NN}(H)$ scales linearly with $N$ for odd-$N$ networks 
is in apparent contradiction with the inset of Fig.~\ref{Nby1}, but 
consistency is recovered once one notes that $R(0)\propto 1/N$ so that, 
at high fields, $R_{NN}(H)/R(0)\propto N^2$ for $N>1$.

Since Eq.~(\ref{univ}) is independent of $M$, we expect it to be valid 
for all uniform networks.
However, in the case of large random networks, the situation is more 
complicated since there is generally a distribution of eigenvalues that 
approaches zero at high fields.  This must be properly treated using 
sophisticated tools such as random matrix theory. 

%
%
\section{Large square networks}\label{large}
To investigate larger networks, we focus on $N\times N$ networks because 
their zero-field resistance remains finite as $N\rightarrow \infty$.
Of course, a finite zero-field resistance is obtained for any 
$N\times M$ network provided we keep the ratio $N/M$ constant as 
$N\rightarrow \infty$, but square networks are chosen for 
numerical convenience.

%
%
\subsection{Uniform resistor networks}\label{uni}
\begin{figure}[htbp]
\begin{center}
\includegraphics[width=7cm]{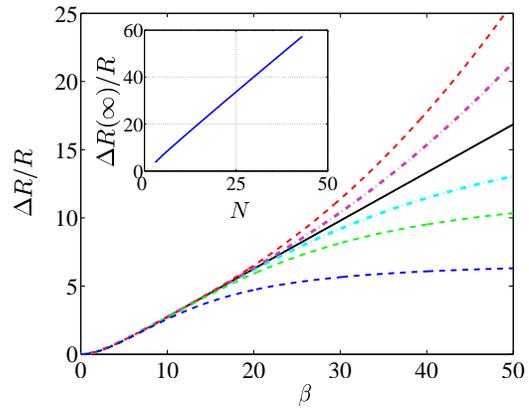}
\caption{(Color online)
  Magnetoresistance of $N\times N$ uniform networks, where the 
  saturating dotted curves correspond to $N =$ 5, 9 and 13, 
  in order of increasing size,  
  and the remaining dotted curves are $N=$ 10 and 14 as curve size decreases.
  The solid curve represents the straight line 
  $\Delta R/R \simeq 0.35 \beta$ that the dotted curves collapse onto when 
  $1< \beta < 2N$.
  Inset: for odd $N$ the saturation level $\Delta R(\infty)/R \simeq 1.4 N$}
\label{MRNN}
\end{center}
\end{figure}

The simplest square network is where all the resistors are identical 
and in this case the zero-field resistance is constant as system size is 
increased.
In accordance with Eq.~(\ref{univ}), uniform square networks retain the 
`odd-even' trend of $N\times 1$ networks, where odd-$N$ networks display a 
saturating magnetoresistance and even-$N$ networks exhibit a non-saturating 
one (see Fig.~\ref{MRNN}).  The key difference is that the $\Delta R/R$ 
curves collapse onto a straight line for $\beta > 1$ as $N\rightarrow \infty$,
while there are no changes to the low field ($\beta < 1$) behaviour, 
where $\Delta R/R \propto \beta^2$. 
This non-saturating, linear behaviour resembles the magnetoresistance of the 
silver chalcogenides and it makes large networks candidates for sensors of 
high magnetic fields.
Note that $\Delta R/R$ is independent of $s$ since it just appears as 
a scaling of $R(H)$ in uniform networks.  From the inset of Fig.~\ref{MRNN}, 
we see that the magnetoresistance saturation of odd networks scales linearly 
with $N$ as expected.

\begin{figure}[htbp]
\begin{center}
\includegraphics[width=7cm]{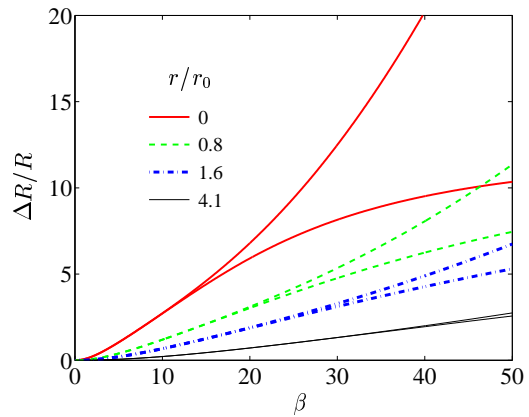}
\caption{(Color online)
  Magnetoresistance of $9\times 9$ and $8\times 8$ uniform networks, 
  where there is a constant resistance $r$ at the connections between 
  elements. Note that $\Delta R/R$ of the $8\times 8$ network 
  corresponds to the larger curve at given $r$. The unit in which $r$ is 
  measured, $r_0$, is taken to be $R(0)$ of a uniform square network with 
  $r=0$.}
\label{MRr}
\end{center}
\end{figure}
%
From the point of view of experimentally constructing $N\times N$ 
uniform networks, it is worth examining the effect of adding a finite 
resistance $r$ at the connections between elements.
One can mathematically show that this is, in fact, equivalent to reducing the 
angular width $\varphi$ of the terminals.
From our numerical simulations, we find that it does not change whether the 
magnetoresistance is saturating or otherwise, but in Fig~\ref{MRr} we see that 
the size of $\Delta R/R$ at a given field decreases with increasing $r$.  
Additionally, it changes the field scale 
such that the divergence of odd and even curves, as well as the 
crossover from linear to quadratic behaviour, is shifted to higher fields.
The reduction in $\Delta R/R$ as $r$ increases is not surprising, because 
in the limit where $r\rightarrow\infty$, the Ohmic dissipation in the network 
is dominated by the connections between disks and we effectively recover a 
network of two-terminal resistors, which has $\Delta R/R = 0$.

\begin{figure}[htbp]
\begin{center}
\includegraphics[width=7cm]{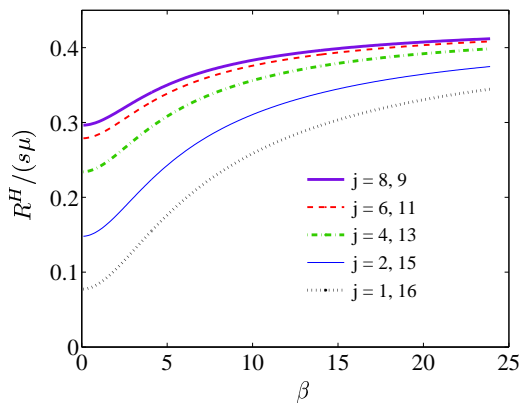}
\caption{(Color online)
  Hall coefficient $R^H_j$ for different positions $j$ across a 
  uniform $16\times 16$ network. It is dependent on $\mu$ as well as $s$, and 
  it has the general form $R^H_j = s\mu f_j(\beta)$.}\label{hall}
\end{center}
\end{figure}

Another important characterization of uniform networks is the Hall coefficient
$R^H$.  In terms of network parameters (see Fig.~\ref{resnet}), we have 
\begin{equation}
  R^H_j = \frac{V^H_j - V^H_{M+j}}{H\sum^N_i I^L_i}
\end{equation}
where it is a function of the position across the network $j = 1, 2 \ldots M$.
By symmetry, we expect $R^H_j = R^H_{M+1-j}$.
The Hall coefficient converges rapidly with increasing system size, so we 
will restrict our consideration to uniform $16\times 16$ networks.
In Fig~\ref{hall}, we observe that it has the general form 
$R^H_j = s \mu f_j(\beta)$ so, if we take $\rho = (ne\mu)^{-1}$, 
$R^H$ is inversely proportional to carrier density $n$
like conventional semiconductors.  In contrast to a conventional 
semiconductor, it is also dependent on $\beta$ at low fields and the 
strength of this dependency increases as we approach the network 
edges $j=1,16$. This already hints that network boundaries play an 
important role in the magnetotransport of uniform networks, as will 
be discussed in Sec.~\ref{discuss}.

%
%
\subsection{Random resistor networks}\label{ran}
\begin{figure}[htbp]
\begin{center}
\includegraphics[width=7cm]{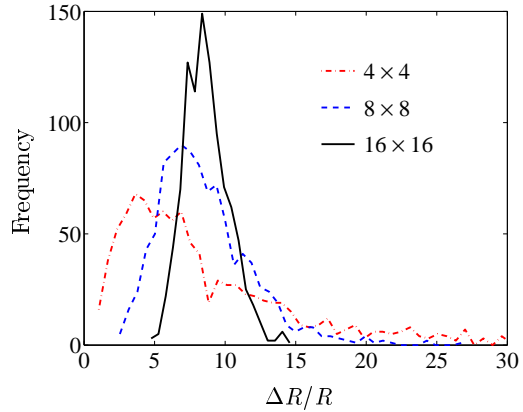}
\caption{(Color online)
  Distributions of magnetoresistance taken from 1000 samples for each 
  network size, where we have set $\langle\mu\rangle = 0$, $\Delta\mu H = 50$, 
  and $\left< \eta \right> = 0$ (so that 
  $\Delta s/\left< s \right> = 1/\sqrt{2}$).
}\label{scale}
\end{center}
\end{figure}

To model real inhomogeneous conductors, it is necessary to consider 
random resistor networks.  In this case, we take the distribution of $\mu$ 
within the network to be Gaussian, with width $\Delta\mu$. 
Since $s$ is always positive, we take $s=\eta^2$, where $\eta$ also has a 
Gaussian distribution of width $\Delta\eta$.  We can then define the 
width of $s$ to be 
$\Delta s = \sqrt{\left< \eta ^4\right> - \left< \eta^2 \right>}$,
where $\left< \cdots \right>$ is an average over the Gaussian distribution.
A numerical analysis of random $N\times N$
networks produces positive magnetoresistances that depend on the 
particular network 
configuration for small $N$ and, consequently, exhibit a large range in 
behaviour whose variation increases with increasing $H$.
However, this range in behaviour diminishes with increasing $N$, 
as illustrated in Figure~\ref{scale}.  The distributions of magnetoresistance 
at large field clearly show a decreasing distribution width as $N$ increases 
and the distribution becomes evenly spread about the mean for 
sufficiently large $N$.
Therefore, the magnetoresistance of the infinite random network should be 
given by the average magnetoresistance of finite networks.

\begin{figure}
\begin{center}
   \includegraphics[width=7cm]{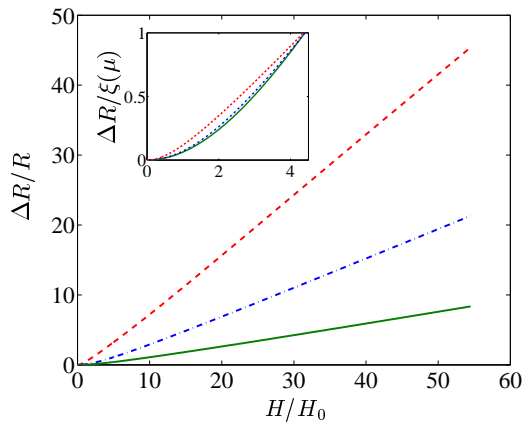}
   \caption{\label{rand}
   (Color online)
   Average magnetoresistance $\Delta R/R$, as a function 
   of dimensionless magnetic field $H/H_0$, of $20\times 20$ random 
   resistor networks for 3 different mobility distributions, where 
   $H_0 = 1$kOe is a typical field scale.  The magnetoresistance was 
   averaged over 10 random network configurations and, in order of 
   increasing size, the curves correspond to:   
   (i) $\langle\mu\rangle = 0.1H_0^{-1}$, $\Delta\mu = H_0^{-1}$, 
   (ii) $\langle\mu\rangle = H_0^{-1}$, $\Delta\mu = 0$, and 
   (iii) $\langle\mu\rangle = 0$, $\Delta\mu = 5H_0^{-1}$. 
   Inset: By scaling the curves so that they all have the same 
   magnetoresistance at 
   around 4$H_0$, it can be seen that linearity continues down to lower 
   fields when the mobility disorder is large, $\Delta\mu \gg H_0^{-1}$}
\end{center}
\end{figure}
Fig.~\ref{rand} displays our key results for simulations performed on 
$20\times 20$ random networks.  We find that the average $\Delta R/R$ is 
linearly dependent on field and that it is strongly dependent on $\mu$, 
but it is still insensitive to $s$ like in the uniform case.
This linear dependence can be argued on the grounds that current in a strongly 
disordered medium at large fields is forced to flow perpendicular to the 
applied voltage a significant proportion of the time and therefore contributes 
the Hall resistance $\rho_{xy}\propto H$ to the effective magnetoresistance. 
The character of the mobility distribution determines the size of the relative 
magnetoresistance because, at sufficiently large magnetic fields, we see that 
$\Delta R/R\propto \langle\mu\rangle$ 
for $\Delta\mu/\langle\mu\rangle < 1$ and $\Delta R/R\propto \Delta\mu$ 
for $\Delta\mu/\langle\mu\rangle > 1$, where the exact proportionality 
constants depend on the details of the distribution.  
Therefore, we would expect $\Delta R/R$ of an 
inhomogeneous semiconductor to diminish with increasing temperature, 
since this corresponds to a decrease in $\mu$ due to phonon excitations.  
This is consistent with experiments on the silver chalcogenides \cite{silver1}.

%

The crossover from linear to quadratic behaviour occurs at field 
$\langle\mu\rangle^{-1}$ for $\Delta\mu/\langle\mu\rangle < 1$  and 
$(\Delta\mu)^{-1}$ for $\Delta\mu/\langle\mu\rangle > 1$.  Thus, even when the 
characteristic field $\langle\mu\rangle^{-1}$ is of order 1T, the measured 
crossover field of a disordered semiconductor can be several orders of 
magnitude smaller, provided $\Delta\mu$ is large.  This yields a 
possible explanation for why the linearity of the silver chalcogenide 
magnetoresistance continues down to fields as low as 10Oe.


It is also of interest to determine how disorder affects the Hall 
coefficient, because experiments on the silver chalcogenides 
have established that an anomalous universal relationship exists 
between the magnetoresistance and the Hall resistance~\cite{silver2}.
Unfortunately, an enormous range in behaviour is displayed for the 
Hall resistance of finite random networks, and there is no obvious 
convergence in the behaviour for network sizes $N<30$.
Thus, we need to examine even larger networks in order to determine the 
Hall resistance for the infinite network.  One possible approach is to 
implement a numerical renormalization group technique where each resistor 
unit is replaced by a new, renormalised resistor unit consisting of a 
$2\times 2$ resistor network, but this is beyond the scope of this paper.

%
%
\section{Boundary Effects}\label{discuss}

\subsection{Effects of macroscopic boundaries}
Before we conclude our study of large resistor networks, we need to  
address an apparent conundrum: 
the uniform network in the infinite limit should behave like a classical 
homogeneous conductor with no magnetoresistance.
To see this, consider a single resistor, like Fig.~\ref{res}, within the 
infinite uniform network. From translational symmetry, current entering the 
resistor from the right (bottom) terminal is equal to the current leaving 
from the left (top).  If we assume that the current flowing perpendicular to 
the applied voltage is zero, as dictated by the boundary conditions, 
then the magnetoresistance of the uniform network becomes equivalent to that 
of a single resistor, and is thus zero.  So why is the magnetoresistance 
that we calculated for the infinite uniform network non-zero and 
non-saturating?
The answer lies in boundary effects due to the perfectly conducting 
electrodes that are used to apply the potential difference across the 
network.

\begin{figure}[htbp]
\begin{center}
\includegraphics[width=4.5cm]{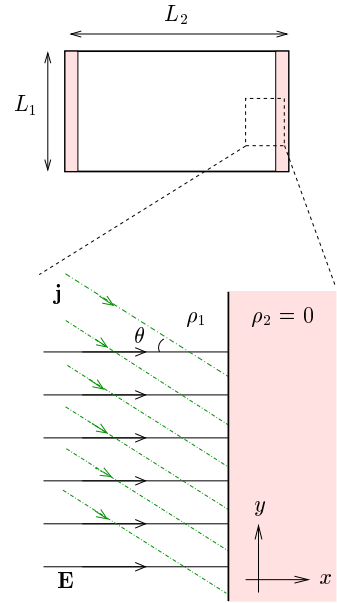}
\caption{(Color online)
  An $L_1 \times L_2$ homogeneous medium with two perfectly 
  conducting electrodes attached. 
  The electrode boundary may be treated as an interface
  between an ideal conductor and a material 
  with finite resistivity $\rho_1$.  In a magnetic field,
  current $\mathbf{j}$ enters the ideal conductor at an angle 
  $\theta = \arctan\beta$ with respect to the electric field $\mathbf{E}$.
  }\label{BEffect}
\end{center}
\end{figure}

Fig.~\ref{BEffect} depicts the boundary between the ideal electrode and 
a material of finite resistivity $\rho_1$ in the $x$-$y$ plane.  
If a magnetic field $H$ is applied in the $z$-direction, the classical 
electrical transport in homogeneous material obeys Ohm's law, with a 
resistivity tensor $\hat\rho$ given by Eq.~(\ref{rho}).
Now, the electric field $E_y$ that is parallel to the 
surface must be continuous across the boundary according to the 
standard Maxwell equations.  
Since the electric field inside an ideal conductor is always zero, then 
$E_y = 0$ and the electric field outside the conductor must therefore be 
perpendicular to the perfectly conducting surface.  
This, combined with the form of $\hat\rho$, causes the current to enter 
and exit the perfectly conducting electrodes at the angle 
$\theta = \arctan \beta$.
For strong fields $\beta \gg 1$, the current is angled at almost  
$90^{\circ}$ to the electric field, so the effective resistivity of the 
material close to the electrodes is 
\begin{equation}
  \rho_{\mathrm{eff}} \ \simeq \ \left | \frac{E_x}{j_y} \right | 
   \simeq \rho_1 \beta
\end{equation}
\begin{figure*}[htbp]
\begin{center}
\includegraphics[height=6.5cm]{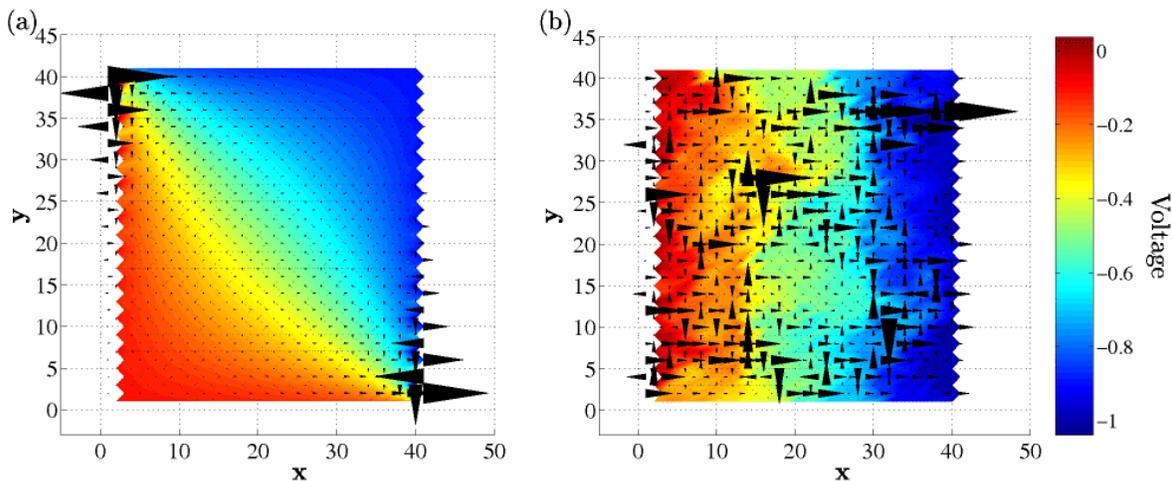}
\caption{(Color online)
  Visualizations of currents and voltages at large field
  in a $20\times 20$ network of disks with radii 1 (arbitrary units), 
  where the potential difference $U = -1$V.  The black arrows 
  represent currents, where arrow size corresponds to the magnitude of 
  the current.
  (a) Uniform network at $\beta =30$.
  (b) Random network with $\langle\mu\rangle = 0$, $\Delta\mu H = 30$ 
      and $\Delta s/\left<s\right> = 1/\sqrt{2}$. 
  }\label{urvisual}
\end{center}
\end{figure*}
This provides an explanation for the linear magnetoresistance of the 
infinite uniform network in Fig.~\ref{MRNN}.  In general, currents will be 
perturbed at a boundary that is perpendicular to the $x$-$y$ plane 
if there is a mismatch of Hall electric fields $E_y$ across the interface 
when $j_y = 0$.


Figure~\ref{urvisual}(a) demonstrates that, in a large uniform network, 
the current is most strongly perturbed at the electrode boundaries in 
a strong magnetic field.  
Deep within the network, away from the boundaries, 
the current is uniformly spread, so a measurement of the bulk 
magnetoresistance using a four-probe\footnote{The effects of 
boundaries on a sample's resistance are 
removed by performing a four-probe measurement, where the voltage probes 
are separate from the electrodes that supply the current, unlike the 
two-probe resistance measurement considered in this paper.}
measurement, will yield a zero magnetoresistance in the infinite size limit.
There is also the extra restriction that no current is allowed to 
enter or leave the top and bottom edges of the network, 
so this forces the majority of the current to enter the network at the 
top left corner and leave at the bottom right corner.
As $\beta \rightarrow \infty$, we obtain singularities of the current 
in the aforementioned corners. 
This type of behaviour has already been noted in the context of 
real homogeneous materials~\cite{pippard}.

In addition, the anomalous Hall coefficient in Fig.~\ref{hall} 
qualitatively matches calculations for the response of simple Hall 
devices constructed from homogeneous materials with a 
square geometry~\cite{W54,P91} like that in Fig.~\ref{BEffect}. 
There, a geometrical correction factor is used to describe the diminution 
of the Hall voltage due to finite size effects, 
and this factor is dependent on the ratio of the electrode width $L_1$ 
with respect to the length $L_2$ of the Hall device,
i.e.\ the ratio $N/M$ of an infinite uniform resistor network.

It is important to stress that \emph{random} resistor networks correspond to 
an entirely different class of system from the homogeneous conductor: 
the infinite random resistor network represents an inhomogeneous material, 
while a uniform network may only be regarded as inhomogeneous when the 
system size is finite.
Comparing Figs.~\ref{urvisual}(a) and (b), we see that the 
current paths are highly inhomogeneous and filamentary within the 
random network unlike the uniform case.  The voltage landscape is also 
nontrivial and the current paths create loops within the random system.  
Therefore, the magnetoresistance should be non-zero deep within the random 
network, away from the boundaries.

We can strengthen this claim by considering the voltage correlation 
function:
\begin{equation}
  V_{corr} (\mathbf{x}) \ = \ \langle 
  V(\mathbf{r+x})-V(\mathbf{r})\rangle_{\mathbf{r}}
\end{equation}
where $V(\mathbf{r})$ is the voltage at position $\mathbf{r}$ in the network, 
$\mathbf{x}$ denotes vectors oriented in the $x$-direction, and 
$\langle \cdots \rangle_{\mathbf{r}}$ represents an average over 
$\mathbf{r}$.
Taking the radii of the disks to be 1 (arbitrary units), 
we should note that $V_{corr}(N\hat x) = U$ in an $N\times N$ network, 
since this corresponds to measuring the potential difference across 
the whole network.

\begin{figure}[htbp]
\begin{center}
\includegraphics[width=7cm]{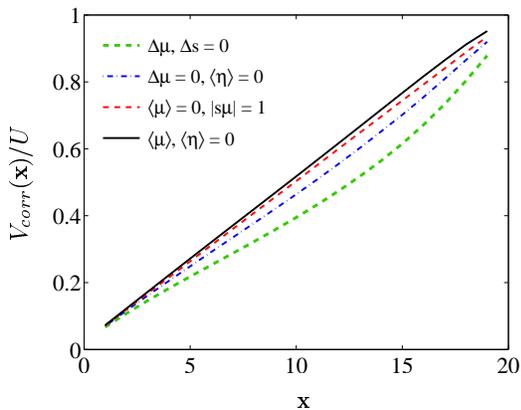}
\caption{(Color online)
  Voltage correlation function at large magnetic field 
  for $20\times 20$ networks of varying disorder.  
  Note that the cases with $\left<\eta\right> = 0$ correspond to 
  $\Delta s = \left<s\right>/\sqrt{2}$, since $s = \eta^2$.
  For random networks, the function has been averaged over 
  100 samples.}\label{Vxcorr}
\end{center}
\end{figure}

Figure~\ref{Vxcorr} plots the voltage correlation function at large magnetic 
field of $20\times 20$
networks for four different cases of disorder.  When the network is uniform, 
we see that the slope of $V_{corr}(\mathbf{x})$ suddenly increases for 
large $\mathbf{x}$, which implies 
that the voltage drops within the network are smaller than those close to 
the boundary.  This demonstrates that much of the magnetoresistance 
of the network is confined to the boundaries, as expected.
In contrast, as we increase the disorder in the network, the 
behaviour of $V_{corr}(\mathbf{x})$ tends to a straight line, indicating 
that the linear magnetoresistance is spread across the whole of the network 
and is not just a boundary effect.  
Of particular interest is the fact that maximum insensitivity to the 
boundary is achieved when $\langle \mu \rangle = 0$, 
which could imply that the magnetoresistance is largest when electrons and 
holes are present in equal proportions, as has been measured in 
experiment~\cite{silver3}.
Recent calculations by Guttal and Stroud~\cite{GS05} on two-dimensional, 
two-phase media further support these observations.  They prove that the 
magnetoresistance of their inhomogeneous conductor is linear when 
$\left<\mu\right> = 0$, but it can saturate when $\left<\mu\right> \neq 0$, 
like in experiment.

\subsection{Finite-element modelling of macroscopic media}

In addition to studies of random media, there is also interest in 
understanding classical macroscopic media with complex geometries, because 
of possible geometrical effects in magnetoresistance and Hall measurements.
Geometric enhancements of the magnetoresistance are already the basis 
of sensitive EMR magnetic-field sensors~\cite{STHH00,ZHS01-2}.
In general, it is difficult to analytically calculate the magnetotransport 
of macroscopic media with complicated boundaries, 
because the calculation involves solving differential equations for the 
currents/voltages where the boundary conditions contain derivatives 
that are oblique to the boundary surfaces~\cite{poincare}.
Thus, standard mathematical techniques, such as 
separation of variables, will typically fail in these problems, although 
the application of conformal mappings in two dimensions has proved successful 
in dealing with simple geometries~\cite{W54,RG81,LA89}.
However, one can, in principle, use infinite uniform networks to simulate 
two-dimensional, macroscopic, composite conductors, in a manner 
analogous to the finite-element modelling of EMR devices~\cite{MRSZHS01}.
Moreover, current perturbations at the connections between disks can be 
disregarded entirely in these networks if we choose the terminals 
to have the same resistivity as the disks.
Boundaries within the macroscopic system still present a potential problem 
since they involve disks of differing resistivity connected together.
But, at fixed magnetic field, the magnitude of these contact effects 
will tend to zero as the granularity of the network goes to zero  
($N,M\rightarrow\infty$), provided the 
number of elements with contact effects scales slower than the total 
number of elements.  This is certainly true for one-dimensional 
boundaries within a two-dimensional homogeneous medium, 
since the number of boundary elements in the network scales 
like $N$ while the total number of elements scales like $N^2$. 

It is important to note that the homogeneous conductor constructed 
from the infinite uniform network will possess a mobility 
$\mu^{*}$ and resistivity $\rho^{*}$ that is different 
from those of the elements that generate the network.  Generally, these 
effective network parameters will depend on the geometry of the element 
as well as the connections between elements.  For the situation where 
there is no resistance between the elements, the effective quantities can 
be determined from the resistivity and Hall coefficient of a single element.
Therefore, using Eq.~(\ref{z_disk}), we have parameters
\begin{eqnarray}
  \rho^{*} & = & \frac{2\rho}{\pi t}(g(\varphi) + 0.35) \\ 
  \label{mueff}
  \mu^{*} & = & \frac{\pi\mu}{2(g(\varphi) + 0.35)}
\end{eqnarray}
The high-field magnetoresistance of the uniform square network in 
Fig.~\ref{MRNN} is then given by $\Delta R/R \sim \mu^{*} H$.

To confirm the validity of our numerical approach, we can compare the 
magnetoresistance of infinite uniform networks 
with results of the conformal mapping approach. 
Following the method of Rendell and Girvin~\cite{RG81}, we find that the 
$L_1\times L_2$ homogeneous medium in Fig.~\ref{BEffect} has resistance:
\begin{equation}
  R \ = \ \rho_1\sqrt{1+\beta^2} \ \frac{\int_0^1 dx \ 
        \cos\Theta_{L_1 L_2}(x)}
       {\int_0^1 dx \ \cos\Theta_{L_2 L_1}(x)}
\end{equation}
where
\begin{align*}
  \Theta_{L_1 L_2}(x) = & 
  \sum_{n (\mathrm{odd})}\frac{4\arctan(\beta)}{n\pi}\frac{\sin(n\pi x)}
      {\cosh(\frac{n\pi L_1}{2L_2})}
\end{align*}
Thus, for the special case where $L_1=L_2$, we have the exact result 
$R = \rho_1\sqrt{1+\beta^2}$, which is in agreement 
with our numerical simulations of $N\times N$ networks in Fig.~\ref{MRNN}
if we take  $N\rightarrow\infty$, $\beta = \mu^{*}H$ and 
$\rho_1 = \rho^{*}$.

\begin{figure}
\begin{center}
   \includegraphics[width=7cm]{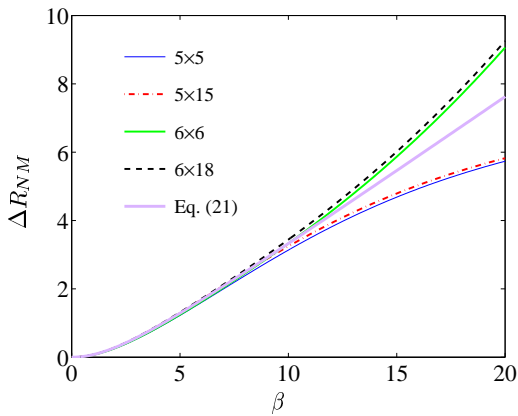}
\end{center}
   \caption{\label{MR_NM}
   (Color online)
   Comparison of resistances from a representative sample of 
   $N\times M$ networks with $M\geq N$, where we have defined 
   $\Delta R_{NM} \equiv R_{NM}(H) - R_{NM}(0)$ and we have set $s=1$.
   These curves approach Eq.~(\ref{infR}) in the limit of infinite 
   network size.
   Curves with $M > 3N$ are indistinguishable from the $M = 3N$ curves 
   and are therefore not shown.}
\end{figure}

For a general $N\times M$ uniform network with $M\geq N$, 
the numerical simulations in Fig.~\ref{MR_NM} demonstrate that the 
resistance is approximately given by the expression:
\begin{equation}
 R_{NM}(H) \ \simeq \ R_{NN}(H) \ + \ R_{NN}(0)\left(\frac{M}{N} - 1 \right)
\end{equation}
%
Thus, in the limit where $N,M\rightarrow\infty$, we have resistance 
\begin{equation}\label{infR}
R_{NM} \simeq \rho^{*}\left(\sqrt{1 + (\mu^{*} H)^2} - 1 + \frac{M}{N} \right)
\end{equation}
This becomes independent of network dimensions as $H \rightarrow \infty$, 
like previous studies of two-terminal devices have predicted~\cite{IK92}, 
but the high-field magnetoresistance is $\Delta R/R \propto N/M$ 
since $R_{NM}(0) = \rho^{*}M/N$.

\subsection{Contact resistance between elements}
The insights gained in the previous section can be used to 
analyse contact effects 
between elements in the network.  We previously assumed a uniform current 
injection into the disk terminals, but this is generally difficult to 
achieve when $\beta \gg 1$.  The assumption of uniform current 
is only valid for all fields when the Hall resistance $\rho\mu$ and 
thickness $t$ is identical for each element.
To assess the ramifications of current perturbations at the disk terminals, 
we consider the simple case of ideal metal bridges connecting the disks.  
Our numerical simulations show that current distortions are restricted 
to the vicinity of the perfectly conducting electrode, while other 
calculations~\cite{RHWGKWAPW88} demonstrate that the lengthscale of the 
current distortion is proportional to $\varphi$.
Therefore, when $\varphi\ll 1$, it is legitimate to replace the 
bridge with a two-terminal resistor possessing a field-dependent resistance.  
Using our results for a homogeneous conductor with boundaries, 
we determine the contact resistance between disks $i$ and $j$ to be
\begin{equation} \label{conMR}
  \rho^{ij}_{c} = \frac{1}{2} \left( \frac{\rho_i}{t_i} h(\mu_i H)
  + \frac{\rho_j}{t_j} h(\mu_j H) \right)
\end{equation}
where we take $h(\beta) = \sqrt{1+\beta^2}-1$ to obtain the correct low-field 
and high-field limits.

The contact resistance cannot be eliminated from our network model by 
reducing the 
terminal width $\varphi$, because the disk impedance coefficients associated 
with field in Eq.~(\ref{z_disk}) are independent of $\varphi$ like 
$\rho_{c}$, even though the disk resistance tends to infinity as 
$\varphi\rightarrow 0$.
Moreover, in the case of the infinite uniform network at large fields, 
the contact resistance $\rho_c$ is equal to the network resistance 
$\Delta R_{NM}$ without metallic bridges for \emph{all} $\varphi\ll 1$.

From simulations of random square networks that include this contact 
resistance, we find that the size of the network magnetoresistance is 
increased by up to 100\% or more, depending on the network disorder, 
but the major results are qualitatively unchanged: the variation in 
magnetoresistance decreases with increasing network size like in 
Fig.~\ref{scale}, the average 
magnetoresistance is linearly dependent on field, and the crossover point is 
determined by the mobility distribution, as in Fig.~\ref{rand}.
One important consequence of these contact resistances is that the bulk 
magnetoresistance is always non-saturating and linear, plus the network's 
sensitivity to the network boundaries is reduced.  For example, an infinite 
uniform square network with metal bridges will possess a non-zero bulk 
magnetoresistance of $\rho_c/R(0)$ due to the contacts, and we find that 
this accounts for about 70\% of a two-terminal measurement of the
network's magnetoresistance.

Note that contacts of perfectly-conducting wires represent an extreme limit 
where the contact effects are greatest.
A potentially richer case is where the wires are replaced by interfaces 
between elements.  Here, the magnitude of each contact resistance 
in a random network ranges from Eq.~(\ref{conMR}) right down to zero, when 
neighbouring elements have the same Hall resistance and thickness.  
Thus, we can expect to recover bulk magnetoresistances, similar to those 
displayed in Fig.~\ref{Vxcorr}, that sensitively depend on network disorder.  

The situation is further complicated when we consider three-dimensional 
effects at the connections between resistors.
An interesting example is where the disk resistivity $\rho$ and mobility $\mu$ 
are constant within the network, but $s$ is varied by altering the disk 
thickness $t$.  Calculations by Bruls et al.~\cite{bruls,BBGKW85} 
that involve mapping sample thickness variations onto a two-dimensional 
problem, have shown that sharp changes in thicknesses, like those at the 
interfaces between resistors, will have resistance
\begin{equation}
  \rho^{ij}_{c} \ \sim \ \rho \beta\Delta_{ij} 
\end{equation}
where $\Delta_{ij}\equiv \frac{t_i-t_j}{t_i}$, $t_i\geq t_j$ and 
$\beta\Delta_{ij} \gg 1$.  Thus, the contact resistance still has 
a linear field dependence at large fields, but the size of the effect 
is reduced so that the assumption of uniform current injection is 
now valid for $\beta \ll \Delta^{-1}_{ij}$.
A thorough analysis of these more complex contact effects will 
be the subject of future work.

%
%
\section{Conclusion}
\label{conc}

In this paper we have modelled an inhomogeneous conductor using a 
two-dimensional random resistor network that consists of \emph{four}-terminal 
resistors in order to take account of the Hall component.
We have shown that the network impedance matrix $Z$ becomes an 
odd, antisymmetric matrix at large magnetic field, so that the high-field 
behaviour of the magnetoresistance is determined by the zero eigenvalue 
of $Z$.
We find that a non-saturating magnetoresistance can be obtained in 
networks as small as $2\times 1$, where a plurality of current paths 
is allowed within the network, while large networks typically possess 
a linear magnetoresistance. 
This is in contrast to EMR devices that exhibit an extremely large 
but saturating magnetoresistance~\cite{STHH00,ZHS01-2}.

By considering large square networks, we have demonstrated that 
uniform networks in the limit of infinite size are equivalent to 
homogeneous conductors and the observed linear magnetoresistance in 
this system results from boundary effects at the macroscopic electrodes.  
As such, they can be used to model macroscopic media with complex boundaries.
However, large random networks model strongly inhomogeneous semiconductors 
and their magnetoresistance is not simply a boundary effect.  
They correctly reproduce the anomalous magnetoresistance of the silver 
chalcogenides: non-saturating behaviour with a linearity that continues 
down to low fields for large mobility disorder. 
Moreover, the magnetoresistance may be large when the Hall resistance is 
zero, like in experiment.

The advantage of such a phenomenological model of positive, non-saturating 
magnetoresistance is that it is potentially relevant to a whole range 
of materials. Already, similar magnetoresistances have been observed in 
metallic VO${}_x$ thin films~\cite{RKKH03}, micro-sized Co$_x$-C$_{1-x}$ 
composites~\cite{XZZ03} and LaSb$_2$ crystals~\cite{YGDGACH03}. 
Two-dimensional electron gases also show a mysterious linear 
magnetoresistance~\cite{RBMWPW89} and classical disorder has 
been cited as a possible cause~\cite{SH94}.

A major limitation of our random resistor network model is that it 
is restricted to two dimensions and, thus, cannot describe longitudinal 
magnetoresistance.  It is also known that weakly disordered systems 
with continuous fluctuations in the conductivity possess magnetoresistances
that depend on the dimensionality~\cite{I92}.  Therefore, we must extend 
our resistor network model to three dimensions in order to fully simulate 
an inhomogeneous semiconductor.  However, we anticipate that our 
two-dimensional resistor networks will motivate experiments on the 
magnetotransport of systems with controlled disorder and on high-field 
magnetic sensors.

Finally, it would be interesting to explore the magnetothermopower of our 
resistor networks.  A giant magnetothermopower is associated with 
Ag$_{2-\delta}$Te samples~\cite{SSLR03} and its origin may also lie 
in classical disorder, because experiments on composite 
semiconductor-metal structures demonstrate that the 
magnetothermopower can be geometrically enhanced~\cite{HTM01}.


\begin{acknowledgments}
We are grateful to Anke Husmann, Tom Rosenbaum and Nigel Cooper for 
simulating discussions.
M. M. P. acknowledges support from the Commonwealth Scholarship Commission 
and the Cambridge Commonwealth Trust.
\end{acknowledgments}

\appendix

\section{Impedance matrix of a 4-terminal circular disk}\label{appA}
We begin by writing the electric field in terms of the potential, 
${\bf E}=-\nabla U$, and then combining charge conservation 
$\nabla.\ {\bf j}=0$ with Ohm's law to obtain the following 
differential equation for the potential:
\begin{align}\label{diffeq}
  \frac{\partial}{\partial x_{i}}
  \left(\sigma_{ik}\frac{\partial U}{\partial x_{k}}\right) & = 0
\end{align}
where the conductivity tensor $\hat\sigma = \hat\rho^{-1}$.
For the case of a homogeneous medium, Eq.~(\ref{diffeq}) is simply
the Laplace equation.

Consider the homogeneous disk in Fig.~\ref{res}.
If we assume uniform current injection into the terminals, then we 
can use the currents as the boundary conditions to solve the 
Laplace equation for the potential.  In the absence of a magnetic 
field, it is sufficient to take $\varphi \ll 1$ in order for 
this assumption to be valid, but the currents will generally be distorted 
when $\beta\gg 1$.  To simplify the problem, we shall initially neglect 
these distortions.

Taking the currents entering each terminal to be $I_{1}$, $I_{2}$, $I_{3}$, 
and $I_{4}$, respectively, we then obtain 
the following potential along the edge of the disk:
\begin{eqnarray} \nonumber
 U(\beta,\theta) & = &
   -\frac{\rho}{\pi\varphi t}
      \sum_{n=1}^{\infty}\frac{1}{n^{2}} 
           [(S - \beta T)\cos(n\theta)  \\ \label{sum}
        & &   +  (T + \beta S)\sin(n\theta)]
\end{eqnarray}
where $\theta$ defines the angular position on the disk edge, 
and we have 
\begin{eqnarray*}
S & = & 2I_{1}\sin(n\varphi/2) \\
& & + \ I_{2}[\sin(n\pi/2+n\varphi/2)-\sin(n\pi/2-n\varphi/2)] \\
& & + \ I_{3}[\sin(n\pi+n\varphi/2)-\sin(n\pi-n\varphi/2)] \\
& & + \ I_{4}[\sin(3n\pi/2+n\varphi/2)-\sin(3n\pi/2-n\varphi/2)] \\
T & = & I_{2}[\cos(n\pi/2-n\varphi/2)-\cos(n\pi/2+n\varphi/2)]\\
& & + \ I_{4}[\cos(3n\pi/2-n\varphi/2)-\cos(3n\pi/2+n\varphi/2)]
\end{eqnarray*}
%


To determine the impedance matrix $z$, we take the potential 
differences between the equally-spaced terminals, i.e. $U(\beta,\pi/2)$-$U(\beta,0)$, 
$U(\beta,\pi)$-$U(\beta,\pi/2)$, $U(\beta,3\pi/2)$-$U(\beta,\pi)$ and 
$U(\beta,0)$-$U(\beta,3\pi/2)$, and then sum up the series in 
Eq.~(\ref{sum}) for a sufficient number of terms.  





\end{document}